\documentclass{article}%
\usepackage{amsmath}
\usepackage{amsfonts}
\usepackage{amssymb}
\usepackage{graphicx}%
\providecommand{\U}[1]{\protect\rule{.1in}{.1in}}
\newtheorem{theorem}{Theorem}
\newtheorem{acknowledgement}[theorem]{Acknowledgement}

\begin{document}
\begin{titlepage}
\vspace{.3cm} \vspace{1cm}
\begin{center}
\baselineskip=16pt \centerline{\Large\bf  Hidden Ghost in Massive gravity } \vspace{2truecm} \centerline{\large\bf Ali H.
Chamseddine$^{1,2}$\ , \ Viatcheslav Mukhanov$^{3,4}$\ \ } \vspace{.5truecm}
\emph{\centerline{$^{1}$Physics Department, American University of Beirut, Lebanon}}
\emph{\centerline{$^{2}$I.H.E.S. F-91440 Bures-sur-Yvette, France}}
\emph{\centerline{$^{3}$Theoretical Physics, Ludwig Maxmillians University,Theresienstr. 37, 80333 Munich, Germany }}
\emph{\centerline{$^{4}$LPT de l'Ecole Normale Superieure, Chaire Blaise Pascal, 24 rue Lhomond, 75231 Paris cedex, France}}
\end{center}
\vspace{2cm}
\begin{center}
{\bf Abstract}
\end{center}
The Hessian's determinant for a version of massive gravity given by an infinite expansion of a square root function of the induced metric, vanishes. We
show that it allows us to eliminate one of four scalar fields used to generate the graviton mass.
This, however, gives rise to the appearance of extra terms in the action with the squared time derivative of the metric, thus signaling that a nonlinear ghost survives.
We demonstrate this phenomenon  considering a simple system with constraint, which is supposed to reduces the number of physical degrees of freedom, however, we explicitly show
how the constraint forces the metric to propagate an extra tachyonic state.
\end{titlepage}

\section{\bigskip Introduction}

\bigskip There is now considerable amount of work on a consistent formulation
of massive gravity \cite{hinter}. The breakdown of diffeomorphism invariance
implies that the two helicities $\pm2$ of the massless graviton would be
joined by four degrees to give six degrees of freedom \cite{boul}. This would
then correspond to five degrees of freedom for the massive graviton in
addition to a ghost degree of freedom for the time-like component of broken
diffeomorphism. For the Fierz-Pauli choice of the mass-term the ghost degree
of freedom decouples in the lowest order \cite{pauli}. This model for massive
gravity can be made consistent breaking diffeomorphism invariance
spontaneously by the use of four scalar fields $\phi^{A}$ which will be
absorbed by the massless graviton \cite{tHooft}, \cite{ChM1} . The nonlinear
Boulware-Deser ghost decouples up to the third order but reappears at higher
orders \cite{ChM2}, \cite{ChM3}. It is claimed that for one family of theories
with the action given in terms of an infinite expansion of a square root
function depending on the induced metric \cite{GR1}, the BD ghost decouples to
all orders \cite{HR}. The Hessian of such action was shown to vanish,
indicating that one degree of freedom might disappear, thus reducing the six
naively expected physical degrees of freedom for the gravity and scalar fields
to five and hence removing the nonlinear ghost to all orders of perturbation
theory \cite{GR2}. In this note we will investigate whether this really
happens or not? The family of actions with vanishing determinant of the
Hessian can be simplified by using auxiliary fields \cite{ChM4} . This allows
us to explicitly compute the Hessian, establish that it has a zero mode, and
then eliminate the corresponding degree of freedom expressing the action in
terms of the physical variables. We will show that, however, the process of
elimination of the non-physical scalar field induces new terms in the action
which depend on the squared time derivatives of the metric, and as a result
ghost degree of freedom re-emerges in a nontrivial way. To illustrate this
phenomenon we consider a simple theory, where the constraint is introduced
with the purpose to reduce the number of degrees of freedom for massive
gravity. We show how in this theory the constraint causes the metric itself to
propagate an additional tachyonic state.

\section{Massive gravity in the simplified formulation}

We consider an equivalent formulation of massive gravity \cite{ChM4} to that
in reference \cite{GR1}, which avoids using an infinite expansion in terms of
the induced metric. The equivalence, however, is only valid on shell. We first
define
\begin{align}
S_{AB}^{\prime} &  =e_{A}^{\mu}\partial_{\mu}\phi_{B},\\
S_{AB} &  =S_{AB}^{\prime}-\eta_{AB},
\end{align}
where the sixteen vierbein fields $e_{A}^{\mu}$ are subject to the ten
conditions
\begin{equation}
g^{\mu\nu}=e_{A}^{\mu}e^{\nu A}\equiv e_{A}^{\mu}e_{A}^{\nu},\label{metric}%
\end{equation}
where $g^{\mu\nu}$ is the inverse metric, which will be imposed through
Lagrange multipliers.

The action is given by \cite{ChM4}
\begin{equation}
I=-\frac{1}{2}%
{\displaystyle\int}
d^{4}x\sqrt{g}R+\frac{m^{2}}{2}%
{\displaystyle\int}
d^{4}x\sqrt{g}\left(  S^{2}-S_{AB}S_{AB}\right)  +\frac{1}{2}%
{\displaystyle\int}
d^{4}x\sqrt{g}\tau_{\mu\nu}\left(  g^{\mu\nu}-e_{A}^{\mu}e^{\nu A}\right)  ,
\label{action}%
\end{equation}
where we use the units in which $8\pi G=1.$ One can also add to this action
the terms \cite{CV}
\begin{equation}%
{\displaystyle\int}
d^{4}x\sqrt{g}\left(  c_{1}\delta_{A^{\prime}B^{\prime}C^{\prime}}^{ABC}%
S_{A}^{A^{\prime}}S_{B}^{B^{\prime}}S_{C}^{C^{\prime}}+c_{2}\delta_{A^{\prime
}B^{\prime}C^{\prime}D^{\prime}}^{ABCD}S_{A}^{A^{\prime}}S_{B}^{B^{\prime}%
}S_{C}^{C^{\prime}}S_{D}^{D^{\prime}}\right)  ,
\end{equation}
that do not influence the nature of the ghost decoupling. However, to simplify
the calculations we will not consider them in our analysis. Variation of the
action with respect to the metric $g^{\mu\nu}$ gives
\begin{equation}
R_{\mu\nu}-\frac{1}{2}g_{\mu\nu}R=\tau_{\mu\nu}^{\prime}-\frac{1}{2}g_{\mu\nu
}\tau^{\prime},
\end{equation}
where
\begin{equation}
\tau_{\mu\nu}^{\prime}=\tau_{\mu\nu}+\frac{1}{2}g_{\mu\nu}\Lambda
,\qquad\Lambda=\frac{m^{2}}{2}\left(  S^{2}-S_{AB}S_{AB}\right)  ,
\end{equation}
or equivalently,
\begin{equation}
\tau_{\mu\nu}^{\prime}=R_{\mu\nu}.
\end{equation}
The$\ $variation with respect to $e_{A}^{\mu}$ leads to
\begin{equation}
\tau_{\mu\nu}e_{A}^{\nu}=-m^{2}\left(  S_{AC}\partial_{\mu}\phi_{C}%
-S\partial_{\mu}\phi_{A}\right)  . \label{auxiliary}%
\end{equation}
Multiplying this equation with $e_{B}^{\mu}$ we have%
\begin{align}
e_{B}^{\mu}\tau_{\mu\nu}e_{A}^{\nu}  &  =-m^{2}\left(  S_{AC}S_{BC}^{\prime
}-SS_{BA}^{\prime}\right) \nonumber\\
&  =-m^{2}\left(  S_{AC}S_{BC}-SS_{BA}+S_{AB}-\eta_{AB}S\right)  .
\end{align}
The symmetry of the tensor $\tau_{\mu\nu}$ implies that the left hand side of
the above equation is symmetric with respect to the indices $A$ and $B$%
\begin{equation}
e_{B}^{\mu}\tau_{\mu\nu}e_{A}^{\nu}=e_{B}^{\nu}\tau_{\mu\nu}e_{A}^{\mu},
\end{equation}
and thus
\begin{equation}
S_{AC}S_{BC}-SS_{BA}+S_{AB}=S_{BC}S_{AC}-SS_{AB}+S_{BA},
\end{equation}
or, equivalently
\begin{equation}
(S+1)\left(  S_{AB}-S_{BA}\right)  =0.
\end{equation}
The possible solution
\begin{equation}
S=-1,
\end{equation}
when combined with equation $\left(  \ref{auxiliary}\right)  $ gives
\begin{equation}
\tau_{\mu\nu}=-m^{2}\left(  \partial_{\mu}\phi_{C}\partial_{\nu}\phi
_{C}\right)  .
\end{equation}
This is inconsistent with the vacuum solution \cite{tHooft}:
\begin{equation}
e_{A}^{\mu}=\delta_{A}^{\mu},\qquad\phi^{A}=x^{A},
\end{equation}
for which $S=4.$ We thus assume that $S\neq-1,$ and hence%
\begin{equation}
S_{AB}=S_{BA}. \label{S}%
\end{equation}
which implies that $S_{AB}^{\prime}$ is also symmetric. Using the identity
\begin{align}
S_{AB}^{\prime}S_{AC}^{\prime}  &  =e_{A}^{\mu}\partial_{\mu}\phi_{B}%
e_{A}^{\nu}\partial_{\nu}\phi_{C}=g^{\mu\nu}\partial_{\mu}\phi_{B}%
\partial_{\nu}\phi_{C}=H_{BC},\\
H_{AB}  &  \equiv\eta_{AB}+\overline{h}_{AB},
\end{align}
we conclude that, since $S_{AB}^{\prime}$ is symmetric,
\begin{align}
S_{AB}  &  =S_{AB}^{\prime}-\eta_{AB}=\sqrt{\eta_{AB}+\overline{h}_{AB}}%
-\eta_{AB}\nonumber\\
&  =\frac{1}{2}\overline{h}_{AB}-\frac{1}{8}\overline{h}_{AC}\overline{h}%
_{CB}+\cdots
\end{align}
Substitution of this expression in the action (\ref{action}) leads on shell
(i.e. after solving the constraint equations (\ref{metric})) and (\ref{S})) to
the infinite expansion given in \cite{GR1}.

\section{\bigskip Vanishing of the Hessian's determinant on shell}

We now compute the Hessian of the terms depending on the scalar fields
$\phi^{A}$ in the action (\ref{action}), defined as
\begin{equation}
\mathcal{H}_{CD}=\frac{\delta^{2}I}{\delta\left(  \partial_{0}\phi^{C}\right)
\delta\left(  \partial_{0}\phi^{D}\right)  },
\end{equation}
taking into account that 16 constraint equations
\begin{align}
g^{\mu\nu} &  =e_{A}^{\mu}e^{\nu A},\\
e_{A}^{\mu}\partial_{\mu}\phi_{B} &  =e_{B}^{\mu}\partial_{\mu}\phi
_{A}.\label{symmetry}%
\end{align}
which allow us to express $e_{A}^{\mu}$ in terms of the fields $g^{\mu\nu}$
and $\partial_{\mu}\phi_{A}.$ First we evaluate the variations
\begin{align}
\frac{\delta S}{\delta\left(  \partial_{0}\phi^{C}\right)  } &  =e_{C}%
^{0}+Y_{AC}^{\mu}\partial_{\mu}\phi_{A}\\
\frac{\delta S_{AB}}{\delta\left(  \partial_{0}\phi^{C}\right)  } &
=e_{A}^{0}\eta_{BC}+Y_{AC}^{\mu}\partial_{\mu}\phi_{B}%
\end{align}
where
\begin{equation}
Y_{AC}^{\mu}=\frac{\delta e_{A}^{\mu}}{\delta\left(  \partial_{0}\phi
^{C}\right)  }.
\end{equation}
and setting
\begin{equation}
\left[  \frac{\delta I}{\delta e_{A}^{\mu}}\right]  _{\mathrm{onshell}}=0
\end{equation}
which gives the equation
\begin{equation}
S_{AB}\partial_{\mu}\phi_{B}=S\partial_{\mu}\phi_{A}%
\end{equation}
After some algebra we obtain for the second variation of the action
\begin{align}
\mathcal{H}_{CD} &  =m^{2}\left[  \left(  Y_{CD}^{0}+Y_{DC}^{0}\right)
S-\left(  Y_{AD}^{o}S_{AC}+Y_{AC}^{0}S_{AD}\right)  +e_{C}^{0}e_{D}^{0}%
-g^{00}\eta_{CD}\right.  \nonumber\\
&  \qquad+\left(  e_{C}^{0}Y_{AD}^{\mu}+e_{D}^{0}Y_{AC}^{\mu}\right)
\partial_{\mu}\phi^{A}-e_{A}^{0}\left(  Y_{AD}^{\mu}\partial_{\mu}\phi
_{C}+Y_{AC}^{\mu}\partial_{\mu}\phi_{D}\right)  \nonumber\\
&  \qquad\quad\left.  -Y_{AD}^{\mu}Y_{AC}^{\nu}\partial_{\mu}\phi_{B}%
\partial_{\nu}\phi_{B}+Y_{BD}^{\mu}Y_{AC}^{\nu}\partial_{\mu}\phi_{B}%
\partial_{\nu}\phi_{A}\right]  \label{Hessian}%
\end{align}
where we have used $g^{00}=e_{C}^{0}e_{C}^{0}.$ One can only determine the
functions $Y_{AB}^{\mu}$ perturbatively. To avoid this inconvenience, we first
establish some properties of these functions, and then use them to simplify
the expression of $\mathcal{H}_{CD}.$ First, we differentiate $g^{\mu\nu}$ to
obtain
\begin{align}
0 &  =\frac{\delta g^{\mu\nu}}{\delta\left(  \partial_{0}\phi^{C}\right)
}=\frac{\delta}{\delta\left(  \partial_{0}\phi^{C}\right)  }\left(  e_{A}%
^{\mu}e_{A}^{\nu}\right)  \nonumber\\
&  =Y_{AC}^{\mu}e_{A}^{\nu}+e_{A}^{\mu}Y_{AC}^{\nu},\label{second}%
\end{align}
and, in particular,
\begin{equation}
e_{A}^{0}Y_{AC}^{0}=0.
\end{equation}
From (\ref{symmetry}) we get the relations
\begin{align}
0 &  =\frac{\delta}{\delta\left(  \partial_{0}\phi^{C}\right)  }\left(
e_{A}^{\mu}\partial_{\mu}\phi_{B}-e_{B}^{\mu}\partial_{\mu}\phi_{A}\right)
\nonumber\\
&  =Y_{AC}^{\mu}\partial_{\mu}\phi_{B}-Y_{BC}^{\mu}\partial_{\mu}\phi
_{A}+e_{A}^{0}\eta_{BC}-e_{B}^{0}\eta_{AC}.
\end{align}
Finally differentiating the identity $S_{AB}^{\prime}S_{AB}^{\prime}=g^{\mu
\nu}\partial_{\mu}\phi_{A}\partial_{\nu}\phi_{A},$ we obtain
\begin{align}
0 &  =\frac{\delta}{\delta\left(  \partial_{0}\phi^{C}\right)  }\left(
S_{AB}^{\prime}S_{AB}^{\prime}-g^{\mu\nu}\partial_{\mu}\phi_{A}\partial_{\nu
}\phi_{A}\right)  \nonumber\\
&  =2\left(  S_{AB}^{\prime}\left(  Y_{AC}^{\mu}\partial_{\mu}\phi_{B}%
+e_{A}^{0}\eta_{BC}\right)  -g^{\mu0}\partial_{\mu}\phi_{C}\right)  .
\end{align}
Because
\begin{equation}
S_{AB}^{\prime}e_{A}^{0}\eta_{BC}=g^{0\mu}\partial_{\mu}\phi_{C},
\end{equation}
this implies that
\begin{equation}
S_{AB}^{\prime}Y_{AC}^{\mu}\partial_{\mu}\phi_{B}=0,
\end{equation}
and the $e_{A}^{\mu}$ equations of motion (\ref{auxiliary}) then give
\begin{align}
0 &  =S_{AB}^{\prime}Y_{AC}^{\mu}\partial_{\mu}\phi_{B}=-\frac{1}{m^{2}}%
\tau_{\mu\nu}e_{A}^{\nu}Y_{AC}^{\mu}+\left(  S+1\right)  \partial_{\mu}%
\phi_{A}Y_{AC}^{\mu}\nonumber\\
&  =\left(  S+1\right)  \partial_{\mu}\phi_{A}Y_{AC}^{\mu},
\end{align}
since
\begin{equation}
\tau_{\mu\nu}e_{A}^{\nu}Y_{AC}^{\mu}=-\tau_{\mu\nu}e_{A}^{\mu}Y_{AC}^{\nu}=0.
\end{equation}
Taking into account that $S\neq-1$ it follows from here that
\begin{equation}
\partial_{\mu}\phi_{A}Y_{AC}^{\,\,\mu}=0.\label{identity}%
\end{equation}
This implies that the last term in (\ref{Hessian}) vanishes. Going back to the
expression for the Hessian, we note that
\begin{equation}
-e_{A}^{0}Y_{AD}^{\mu}\partial_{\mu}\phi_{C}=e_{A}^{\mu}Y_{AD}^{0}%
\partial_{\mu}\phi_{C}=S_{AC}^{\prime}Y_{AD}^{0},
\end{equation}
The term before the last in (\ref{Hessian} ) is tricky. To simplify it we
first vary equation (\ref{identity}) with respect to $\delta\left(
\partial_{0}\phi^{C}\right)  $ to get
\begin{equation}
Y_{DC}^{0}+\partial_{\mu}\phi_{A}X_{ADC}^{\mu}=0\label{Y0}%
\end{equation}
where we have defined
\begin{equation}
X_{ADC}^{\mu}=\frac{\delta^{2}e_{A}^{\mu}}{\delta\left(  \partial_{0}\phi
^{C}\right)  \delta\left(  \partial_{0}\phi^{D}\right)  }\label{X}%
\end{equation}
It is clear from the definition of $X_{ACD}^{\mu}$ in equation (\ref{X}) and
the property of $Y_{CD}^{0}$ in equation (\ref{Y0}) that
\begin{equation}
Y_{CD}^{0}=Y_{DC}^{0}.
\end{equation}
Next we vary equation (\ref{second}) to get
\begin{equation}
Y_{AD}^{\mu}Y_{AC}^{\nu}+Y_{AC}^{\mu}Y_{AD}^{\nu}+e_{A}^{\mu}X_{ACD}^{\nu
}+e_{A}^{\nu}X_{ACD}^{\mu}=0
\end{equation}
which implies the identity
\begin{align}
Y_{AD}^{\mu}Y_{AC}^{\nu}\partial_{\mu}\phi_{B}\partial_{\nu}\phi_{B} &
=-e_{A}^{\mu}X_{ACD}^{\nu}\partial_{\mu}\phi_{B}\partial_{\nu}\phi_{B}\\
&  =-S_{AB}^{\prime}X_{ACD}^{\nu}\partial_{\nu}\phi_{B}\\
&  =(S+1)Y_{CD}^{0}%
\end{align}
Thus finally the Hessian simplifies to
\begin{equation}
\mathcal{H}_{CD}=m^{2}\left(  Y_{CD}^{0}\left(  S+1\right)  +e_{C}^{0}%
e_{D}^{0}-g^{00}\eta_{CD}\right)  .
\end{equation}
One can immediately see that $\mathcal{H}_{CD}$ has the null eigenvector
$e_{D}^{0},$%
\begin{equation}
\mathcal{H}_{CD}e_{D}^{0}=0.
\end{equation}
In addition $\mathcal{H}_{CD}$ satisfy the property
\begin{equation}
e_{C}^{i}\mathcal{H}_{CD}e_{D}^{j}=m^{2}\left(  Y^{ij}\left(  S+1\right)
+g^{i0}g^{j0}-g^{00}g^{ij}\right)  ,
\end{equation}
where
\begin{equation}
Y^{ij}=e_{C}^{i}Y_{CD}^{0}e_{D}^{j}=Y^{ji}.
\end{equation}
To find the other eigenvectors we write
\begin{equation}
e_{C}^{i}Y_{CD}^{\,0}=\alpha^{i}e_{D}^{0}-\beta g^{00}e_{D}^{i}.
\end{equation}
From $e_{C}^{i}Y_{CD}^{\,0}e_{D}^{0}=0$ it follows that
\begin{equation}
\alpha^{i}=\beta g^{0i},
\end{equation}
and hence
\begin{equation}
Y^{ij}=\beta\left(  g^{i0}g^{j0}-g^{00}g^{ij}\right)  .
\end{equation}
Therefore we have
\begin{equation}
(g^{00}e_{C}^{i}-g^{0i}e_{C}^{0})Y_{CD}^{0}=-\beta g^{00}(g^{00}e_{D}%
^{i}-g^{0i}e_{D}^{0}),
\end{equation}
and the three eigenvectors $(g^{00}e_{C}^{i}-g^{0i}e_{C}^{0})$ has the same
eigenvalue $-\beta g^{00}.$ To find $\beta$ we write
\begin{equation}
e_{C}^{\mu}Y_{CD}^{0}e_{D}^{\nu}=\beta\left(  g^{\mu0}g^{\nu0}-g^{00}g^{\mu
\nu}\right)  ,
\end{equation}
and after contracting with the inverse metric $g_{\mu\nu}$ one obtains
\begin{equation}
\eta^{CD}Y_{CD}^{0}=-3\beta g^{00}\equiv Y=\frac{\delta e_{C}^{0}}%
{\delta\left(  \partial_{0}\phi_{C}\right)  }.
\end{equation}
Thus we have \
\begin{align}
\mathcal{H}^{\mu\nu} &  \equiv e_{C}^{\mu}\mathcal{H}_{CD}e_{D}^{\nu}\\
&  =\mathcal{H}\left(  g^{\mu0}g^{\nu0}-g^{00}g^{\mu\nu}\right)
\end{align}
where
\begin{equation}
\mathcal{H}=\beta\left(  S+1\right)  +1.
\end{equation}
The terms with second time derivative of the fields $\phi^{A}$ has the form
\begin{equation}
\frac{1}{2}\partial_{0}\phi^{C}\mathcal{H}_{CD}\partial_{0}\phi^{D}%
.\label{expansion}%
\end{equation}
Let us define
\begin{equation}
\phi_{C}=x_{C}+\psi_{\mu}e_{C}^{\mu},
\end{equation}
which can be inverted to determine $\psi_{\mu}$ in terms of the metric
$g^{\mu\nu}$ and the scalars $\phi^{A}:$%
\begin{equation}
\psi_{\mu}=e_{\mu}^{C}\left(  \phi_{C}-x_{C}\right)  ,
\end{equation}
where $e_{\mu}^{C}$ is the inverse of $e_{C}^{\mu}$. Then the term
(\ref{expansion}) becomes
\begin{align}
&  \frac{1}{2}\left(  \mathcal{H}_{00}+2\mathcal{H}_{0D}e_{D}^{i}\partial
_{0}\psi_{i}+\partial_{0}\psi_{\mu}e_{C}^{\mu}\mathcal{H}_{CD}e_{D}^{\nu
}\partial_{0}\psi_{\nu}\right.  \nonumber\\
&  \qquad\left.  +2\psi_{\mu}\partial_{0}e_{C}^{\mu}\mathcal{H}_{CD}e_{D}%
^{\nu}\partial_{0}\psi_{\nu}+\psi_{\mu}\partial_{0}e_{C}^{\mu}\mathcal{H}%
_{CD}\partial_{0}e_{D}^{\nu}\psi_{\nu}\right)
\end{align}
Since the projection of $e_{C}^{\mu}\mathcal{H}_{CD}e_{D}^{\nu}$ is zero if
either $\mu=0$ or $\nu=0,$ the above expression reduces to
\begin{equation}
\frac{1}{2}m^{2}\partial_{0}\psi_{i}\mathcal{H}^{ij}\partial_{0}\psi
_{j}+\cdots
\end{equation}
There are, however, couplings of $\psi_{0}$ to the derivatives of
$\partial_{0}e_{C}^{0}$, e.g.
\begin{equation}
\frac{1}{2}\left(  m^{2}\psi_{0}\psi_{0}\partial_{0}e_{C}^{0}\partial_{0}%
e_{D}^{0}\left(  Y_{CD}^{0}\left(  S+1\right)  +e_{C}^{0}e_{D}^{0}-g^{00}%
\eta_{CD}\right)  \right)
\end{equation}
Note that
\begin{equation}
\partial_{0}e_{C}^{0}e_{C}^{0}=\frac{1}{2}\partial_{0}g^{00}%
\end{equation}
and, hence, the squared time derivatives of the metric appear signalling that
the metric itself can propagate extra degree of freedom. On the other hand the
term $\partial_{0}e_{C}^{0}\partial_{0}e_{C}^{0}$ cannot be expressed only in
terms of the metric, but will also involve the antisymmetric part of
$e_{C}^{0}$.

Moreover the action contains also the terms linear in $\left(  \partial
_{0}\phi_{C}\right)  $ which do not contribute to the Hessian. For example,
after integrating by parts we obtain the following linear term
\begin{equation}
3m^{2}\partial_{0}\left(  \sqrt{g}e_{C}^{0}\right)  e_{C}^{0}\psi_{0},
\end{equation}
which after eliminating the $\psi_{0}$ as auxiliary field gives complicated
contributions which depend on the time derivatives of the metric$.$ We
conclude that although the determinant of the Hessian vanishes, eliminating
the zero mode state $\psi_{0}$ introduces squared time derivatives of the
metric signalling the propagation of the ghost states.

\section{Hamiltonian analysis off shell}

Having proved the equivalence of the on shell formulation of massive gravity
to the square root action, one should be able to see that the metric acquires
second time derivatives without having to go on shell. We again start with the
action $\left(  \ref{action}\right)  $ but now treat $e_{A}^{\mu}$ as
independent variable. To do the Hamiltonian analysis, we have to solve for the
equations for conjugated momenta in terms of the fields $\phi_{A}.$ From the
previous analysis, it is clear that it is more convenient to use the variables
$\psi_{\mu},$ defined via,%
\begin{equation}
\phi_{A}=x_{A}+e_{A}^{\mu}\psi_{\mu},
\end{equation}
instead of $\phi_{A}.$ The part of the action containing scalars is then
\begin{align}
&  \frac{m^{2}}{2}\sqrt{g}\left[  -\left(  g^{\mu\nu}g^{\kappa\lambda}%
\partial_{\mu}\psi_{\kappa}\partial_{\nu}\psi_{\lambda}+2g^{\mu\nu}%
e_{A}^{\kappa}\partial_{\nu}e_{A}^{\lambda}\partial_{\mu}\psi_{\kappa}%
\psi_{\lambda}+g^{\mu\nu}\partial_{\mu}e_{A}^{\kappa}\partial_{\nu}%
e_{A}^{\lambda}\psi_{\kappa}\psi_{\lambda}\right)  \right. \nonumber\\
&  \qquad+\left(  g^{\mu\nu}\partial_{\mu}\psi_{\nu}+e_{A}^{\mu}\partial_{\mu
}e_{A}^{\nu}\psi_{\nu}\right)  ^{2}-6\left(  g^{\mu\nu}\partial_{\mu}\psi
_{\nu}+e_{A}^{\mu}\partial_{\mu}e_{A}^{\nu}\psi_{\nu}\right) \nonumber\\
&  \qquad-2g^{\mu\nu}\delta_{\mu}^{A}\left(  \partial_{\nu}e_{A}^{\kappa}%
\psi_{\kappa}+e_{A}^{\kappa}\partial_{\nu}\psi_{\kappa}\right)  +2e_{A}%
^{\kappa}\delta_{\kappa}^{A}\left(  g^{\mu\nu}\partial_{\mu}\psi_{\nu}%
+e_{A}^{\mu}\partial_{\mu}e_{A}^{\nu}\psi_{\nu}\right) \nonumber\\
&  \qquad\left.  +e_{A}^{\mu}e_{B}^{\nu}\left(  \delta_{\mu}^{A}\delta_{\nu
}^{B}-\eta^{AB}\eta_{\mu\nu}\right)  -6e_{A}^{\mu}\delta_{\mu}^{A}+12\right]
.
\end{align}
The term dependent on the second derivative of $\psi_{\mu}$ is
\begin{equation}
\left(  g^{\mu\kappa}g^{\nu\lambda}-g^{\mu\nu}g^{\kappa\lambda}\right)
\partial_{\mu}\psi_{\kappa}\partial_{\nu}\psi_{\lambda},
\end{equation}
where $\left(  \partial_{0}\psi_{0}\right)  ^{2}$ drops out. The problem now
is that there remains a term linear in $\partial_{0}\psi_{0}$ which is
\begin{equation}
m^{2}\sqrt{g}\partial_{0}\psi_{0}\left[  -3g^{00}+\left(  g^{00}e_{A}%
^{i}-g^{0i}e_{A}^{0}\right)  \left(  \partial_{i}e_{A}^{0}\psi_{0}%
+\partial_{i}e_{A}^{j}\psi_{j}\right)  +\left(  e_{A}^{i}g^{00}-g^{i0}%
e_{A}^{0}\right)  \delta_{i}^{A}\right]  .
\end{equation}
It is clear that the conjugated momentum for the field $\psi_{0}$ does not
allow to solve for $\partial_{0}\psi_{0}$, and therefore we first integrate
the time derivative of $\psi_{0}$ by parts%
\begin{align}
&  -m^{2}\psi_{0}\left[  \psi_{j}\partial_{0}\left(  \sqrt{g}\left(
g^{00}e_{A}^{i}-g^{0i}e_{A}^{0}\right)  \partial_{i}e_{A}^{j}\right)
+\sqrt{g}\partial_{0}\psi_{j}\left(  g^{00}e_{A}^{i}-g^{0i}e_{A}^{0}\right)
\partial_{i}e_{A}^{j}\right. \nonumber\\
&  \qquad\qquad+\frac{1}{2}\psi_{0}\partial_{0}\left(  \sqrt{g}\left(
g^{00}e_{A}^{i}-g^{0i}e_{A}^{0}\right)  \partial_{i}e_{A}^{0}\right)
\nonumber\\
&  \qquad\qquad\left.  -3\partial_{0}\left(  \sqrt{g}g^{00}\right)
+\partial_{0}\left(  \sqrt{g}\left(  e_{A}^{i}g^{00}-g^{i0}e_{A}^{0}\right)
\delta_{i}^{A}\right)  \right]  \label{parts}%
\end{align}
One can now solve the equations
\begin{align}
\frac{1}{m^{2}\sqrt{g}}\pi^{i}  &  =\left(  \left(  g^{0i}g^{0j}-g^{00}%
g^{ij}\right)  \partial_{0}\psi_{j}+\left(  g^{0i}g^{jk}-g^{0j}g^{ik}\right)
\partial_{j}\psi_{k}\right. \nonumber\\
&  +\left(  g^{0j}e_{A}^{0}-g^{00}e_{A}^{j}\right)  \left(  \partial_{j}%
e_{A}^{i}\psi_{0}+\delta_{j}^{i}\partial_{0}e_{A}^{\nu}\psi_{\nu}\right)
\nonumber\\
&  \left.  +\left(  g^{0i}e_{A}^{j}-g^{0j}e_{A}^{i}\right)  \partial_{j}%
e_{A}^{\nu}\psi_{\nu}+\left(  e_{A}^{\mu}g^{i0}-e_{A}^{i}g^{\mu0}\right)
\delta_{\mu}^{A}-3g^{0i}\right)  ,
\end{align}
to express $\partial_{0}\psi_{i}$ in terms of the conjugated momentum $\pi
_{i}.$ The action then becomes
\begin{equation}
\frac{1}{2}m^{2}\sqrt{g}\left[  \frac{1}{m^{4}g}\pi^{i}M_{ij}^{-1}\pi
^{j}-B^{i}M_{ij}^{-1}B^{j}\right]  ,
\end{equation}
where
\begin{align}
M^{ij}  &  =\left(  g^{0i}g^{0j}-g^{00}g^{ij}\right) \\
B^{i}  &  =\left(  g^{0i}g^{kl}-g^{0k}g^{il}\right)  \partial_{k}\psi
_{l}+\left(  g^{0k}e_{A}^{0}-g^{00}e_{A}^{k}\right)  \left(  \partial_{k}%
e_{A}^{i}\psi_{0}+\delta_{k}^{i}\partial_{0}e_{A}^{\nu}\psi_{\nu}\right)
\nonumber\\
&  +\left(  g^{0i}e_{A}^{k}-g^{0k}e_{A}^{i}\right)  \partial_{k}e_{A}^{\nu
}\psi_{\nu}+\left(  e_{A}^{\mu}g^{i0}-e_{A}^{i}g^{\mu0}\right)  \delta_{\mu
}^{A}-3g^{0i}.
\end{align}
The $\psi_{0}$ dependent terms take the form
\begin{equation}
\frac{1}{2}m^{2}\sqrt{g}\left(  \psi_{0}\mathcal{A}\psi_{0}+\mathcal{B}%
\psi_{0}\right)  ,
\end{equation}
where
\begin{align}
\mathcal{A}  &  =-\frac{1}{\sqrt{g}}\partial_{i}\left(  \sqrt{g}\left(
g^{0i}g^{0j}-g^{00}g^{ij}\right)  \partial_{j}\right)  -\frac{1}{\sqrt{g}%
}\partial_{0}\left(  \sqrt{g}\left(  g^{00}e_{A}^{i}-g^{0i}e_{A}^{0}\right)
\partial_{i}e_{A}^{0}\right) \nonumber\\
&  +\left(  e_{A}^{\mu}\partial_{\mu}e_{A}^{0}\right)  ^{2}+2\left(
g^{0i}e_{A}^{j}-g^{ji}e_{A}^{0}\right)  \partial_{j}e_{A}^{0}\partial
_{i}-g^{\mu\nu}\partial_{\mu}e_{A}^{0}\partial_{\nu}e_{A}^{0}-C^{i}M_{ij}%
^{-1}C^{j},
\end{align}
is a second order differential operator with respect to the spacial
coordinates, and we defined
\begin{equation}
C^{i}=\left(  g^{0k}e_{A}^{0}-g^{00}e_{A}^{k}\right)  \left(  \partial
_{k}e_{A}^{i}+\delta_{k}^{i}\partial_{0}e_{A}^{0}\right)  +\left(  g^{0i}%
e_{A}^{k}-g^{0k}e_{A}^{i}\right)  \partial_{k}e_{A}^{0}.
\end{equation}
Similarly
\begin{align}
\mathcal{B}  &  =2\left(  e_{A}^{\mu}\partial_{\mu}e_{A}^{0}\right)  \left(
e_{B}^{\nu}\partial_{\nu}e_{B}^{i}\right)  \psi_{i}-\frac{2}{\sqrt{g}}%
\partial_{i}\left(  \sqrt{g}g^{0i}e_{A}^{\mu}\partial_{\mu}e_{A}^{j}\psi
_{j}\right)  +2g^{ij}\left(  e_{A}^{\mu}\partial_{\mu}e_{A}^{0}\right)
\partial_{i}\psi_{j}\nonumber\\
&  +\frac{2}{\sqrt{g}}\partial_{i}\left(  \sqrt{g}\left(  g^{ij}g^{0k}%
-g^{kj}g^{0i}\right)  \partial_{j}\psi_{k}\right)  +\frac{2}{\sqrt{g}}%
\partial_{i}\left(  \sqrt{g}g^{\mu i}e_{A}^{0}\partial_{\mu}e_{A}^{j}\psi
_{j}\right)  -2g^{\mu i}e_{A}^{j}\partial_{\mu}e_{A}^{0}\partial_{i}\psi
_{j}\nonumber\\
&  -2g^{\mu\nu}\partial_{\mu}e_{A}^{0}\partial_{\nu}e_{A}^{i}\psi_{i}+\frac
{6}{\sqrt{g}}\partial_{i}\left(  \sqrt{g}g^{0i}\right)  -6e_{A}^{\mu}%
\partial_{\mu}e_{A}^{0}+2\delta_{\mu}^{A}\left(  e_{A}^{\mu}e_{B}^{\nu}%
-g^{\mu\nu}\eta_{AB}\right)  \partial_{\nu}e_{B}^{0}\nonumber\\
&  -2\left(  g^{0k}e_{A}^{0}-g^{00}e_{A}^{k}\right)  \left(  \partial_{k}%
e_{A}^{i}+\delta_{k}^{i}\partial_{0}e_{A}^{0}\right)  M_{ij}^{-1}\overline
{B}^{j}-\frac{2}{\sqrt{g}}\delta_{j}^{A}\partial_{i}\left(  \sqrt{g}\left(
e_{A}^{j}g^{0i}-e_{A}^{0}g^{ji}\right)  \right) \nonumber\\
&  +\frac{6}{\sqrt{g}}\partial_{0}\left(  \sqrt{g}g^{00}\right)  -\frac
{2}{\sqrt{g}}\partial_{0}\left(  \sqrt{g}\left(  e_{A}^{i}g^{00}-e_{A}%
^{0}g^{i0}\right)  \delta_{i}^{A}\right) \nonumber\\
&  -\frac{2}{\sqrt{g}}\psi_{j}\partial_{0}\left(  \sqrt{g}\left(  g^{00}%
e_{A}^{i}-g^{0i}e_{A}^{0}\right)  \partial_{i}e_{A}^{j}\right)  ,
\end{align}
where
\begin{align}
\overline{B}^{i}  &  =\left(  g^{0i}g^{kl}-g^{0k}g^{il}\right)  \partial
_{k}\psi_{l}+\left(  g^{0i}e_{A}^{0}-g^{00}e_{A}^{i}\right)  \partial_{0}%
e_{A}^{j}\psi_{j}\nonumber\\
&  +\left(  g^{0i}e_{A}^{k}-g^{0k}e_{A}^{i}\right)  \partial_{k}e_{A}^{j}%
\psi_{j}+\left(  e_{A}^{\mu}g^{0i}-e_{A}^{i}g^{0\mu}\right)  \delta_{\mu}%
^{A}-3g^{0i}.
\end{align}
All remaining terms would depend only on $\psi_{i}$ and their spacial
derivatives and the momenta $\pi^{i}.$ Eliminating the $\psi_{0}$ results to
\begin{equation}
-\frac{1}{8}m^{2}\sqrt{g}\mathcal{BA}^{-1}\mathcal{B}%
\end{equation}
and the full action becomes%
\begin{align}
&  \frac{m^{2}}{2}\sqrt{g}\left[  \left(  g^{ij}g^{kl}-g^{ik}g^{jl}\right)
\partial_{i}\psi_{j}\partial_{k}\psi_{l}+\left(  e_{A}^{\mu}\partial_{\mu
}e_{A}^{i}\right)  \left(  e_{B}^{\nu}\partial_{\nu}e_{B}^{j}\right)  \psi
_{i}\psi_{j}\right. \nonumber\\
&  \qquad\qquad+2\left(  g^{ij}e_{A}^{\mu}-g^{i\mu}e_{A}^{j}\right)
\partial_{\mu}e_{A}^{k}\partial_{i}\psi_{j}\psi_{k}-g^{\mu\nu}\partial_{\mu
}e_{A}^{i}\partial_{\nu}e_{A}^{j}\psi_{i}\psi_{j}\ \nonumber\\
&  \qquad+e_{A}^{\mu}e_{B}^{\nu}\left(  \delta_{\mu}^{A}\delta_{\nu}^{B}%
-\eta^{AB}\eta_{\mu\nu}\right)  -6e_{A}^{\mu}\delta_{\mu}^{A}+12\nonumber\\
&  \qquad+2\delta_{\mu}^{A}\left(  e_{A}^{\mu}e_{B}^{\nu}-g^{\mu\nu}\eta
_{AB}\right)  \partial_{\nu}e_{B}^{i}\psi_{i}+2\delta_{\mu}^{A}\left(
e_{A}^{\mu}g^{ji}-e_{A}^{j}g^{\mu i}\right)  \partial_{i}\psi_{j}\nonumber\\
&  \qquad\qquad\left.  -6g^{ij}\partial_{i}\psi_{j}-6e_{A}^{\mu}\partial_{\mu
}e_{A}^{i}\psi_{i}+\frac{1}{m^{4}g}\pi^{i}M_{ij}^{-1}\pi^{j}-\overline{B}%
^{i}M_{ij}^{-1}\overline{B}^{j}-\frac{1}{4}\mathcal{BA}^{-1}\mathcal{B}%
\right]  ,
\end{align}
where%
\begin{equation}
\left(  M^{-1}\right)  _{ij}=\frac{1}{g^{00}}\left(  g_{ij}^{-1}+\frac
{1}{g^{00}-g^{0k}g^{0l}g_{kl}^{-1}}g^{0k}g^{0l}g_{ik}^{-1}g_{jl}^{-1}\right)
,
\end{equation}
and $g_{ij}^{-1}$ is the inverse of the three dimensional metric $g^{ij}$,
$g^{ik}g_{kj}^{-1}=\delta_{j}^{i}.$

We can easily see the metric dependent part of this action by setting
$\psi_{i}=0:$%
\begin{equation}
\frac{m^{2}}{2}\sqrt{g}\left(  e_{A}^{\mu}e_{B}^{\nu}\left(  \delta_{\mu}%
^{A}\delta_{\nu}^{B}-\eta^{AB}\eta_{\mu\nu}\right)  -6e_{A}^{\mu}\delta_{\mu
}^{A}+12-\overline{B}_{0}^{i}M_{ij}^{-1}\overline{B}_{0}^{j}-\frac{1}%
{4}\mathcal{B}_{0}\mathcal{A}^{-1}\mathcal{B}_{0}\right)  ,
\end{equation}
where
\begin{align}
\mathcal{B}_{0}  &  =\frac{6}{\sqrt{g}}\partial_{i}\left(  \sqrt{g}%
g^{0i}\right)  -6e_{A}^{\mu}\partial_{\mu}e_{A}^{0}+2\delta_{\mu}^{A}\left(
e_{A}^{\mu}e_{B}^{\nu}-g^{\mu\nu}\eta_{AB}\right)  \partial_{\nu}e_{B}%
^{0}\nonumber\\
&  -2\left(  g^{0k}e_{A}^{0}-g^{00}e_{A}^{k}\right)  \left(  \partial_{k}%
e_{A}^{i}+\delta_{k}^{i}\partial_{0}e_{A}^{0}\right)  M_{ij}^{-1}\overline
{B}_{0}^{j}-\frac{2}{\sqrt{g}}\delta_{j}^{A}\partial_{i}\left(  \sqrt
{g}\left(  e_{A}^{j}g^{0i}-e_{A}^{0}g^{ji}\right)  \right) \nonumber\\
&  -\frac{2}{\sqrt{g}}\delta_{j}^{A}\partial_{i}\left(  \sqrt{g}\left(
e_{A}^{j}g^{0i}-e_{A}^{0}g^{ji}\right)  \right)  +\frac{6}{\sqrt{g}}%
\partial_{0}\left(  \sqrt{g}g^{00}\right)  -\frac{2}{\sqrt{g}}\partial
_{0}\left(  \sqrt{g}\left(  e_{A}^{i}g^{00}-e_{A}^{0}g^{i0}\right)  \delta
_{i}^{A}\right)  ,
\end{align}
and
\begin{equation}
\overline{B}_{0}^{i}=\left(  e_{A}^{\mu}g^{0i}-e_{A}^{i}g^{0\mu}\right)
\delta_{\mu}^{A}-3g^{0i}.
\end{equation}
The action is non-local because of the operator $\mathcal{A}^{-1}.$ We can
simplify it further by assuming that $e_{A}^{\mu}$ depend only on time. Then
we have
\begin{align}
\mathcal{A}  &  =\left(  e_{A}^{0}\partial_{0}e_{A}^{0}\right)  ^{2}%
-g^{00}\partial_{0}e_{A}^{0}\partial_{0}e_{A}^{0}-\left(  g^{0i}e_{A}%
^{0}-g^{00}e_{A}^{i}\right)  M_{ij}^{-1}\left(  g^{0j}e_{A}^{0}-g^{00}%
e_{A}^{j}\right) \nonumber\\
&  =\frac{1}{4}\left(  \partial_{0}g^{00}\right)  ^{2}-g^{00}\partial_{0}%
e_{A}^{0}\partial_{0}e_{A}^{0}+3g^{00},
\end{align}
and
\begin{align}
\mathcal{B}_{0}  &  =-3\partial_{0}g^{00}+\delta_{\mu}^{A}\left(  e_{A}^{\mu
}\partial_{0}g^{00}-2g^{\mu0}\partial_{0}e_{A}^{0}\right)  +\frac{6}{\sqrt{g}%
}\partial_{0}\left(  \sqrt{g}g^{00}\right) \nonumber\\
&  -\left(  g^{0i}\partial_{0}g^{00}-2g^{00}e_{A}^{i}\partial_{0}e_{A}%
^{0}\right)  M_{ij}^{-1}\left(  \left(  e_{B}^{\mu}g^{0j}-e_{B}^{j}g^{0\mu
}\right)  \delta_{\mu}^{B}-3g^{0j}\right) \nonumber\\
&  -\frac{2}{\sqrt{g}}\partial_{0}\left(  \sqrt{g}\left(  e_{A}^{i}%
g^{00}-e_{A}^{0}g^{i\,0}\right)  \delta_{i}^{A}\right)  .
\end{align}
It is clear that in the linearized approximation, where $g^{00}=1+h^{00}$, the
field $h^{00}$ acquires terms with second time derivatives, and thus become
ghost like.

The conclusion is that, the determinant of the Hessian indeed vanishes,
indicating that one of the scalar fields, in this case $\psi_{0},$ could be
eliminated completely, but this turns out to be a curse rather than a
blessing: new contributions to the metric appear which force the ghost modes
in gravity, previously protected by diffeomorphism invariance, to propagate.

\section{Constraints and induced degrees of freedom}

We now illustrate the conclusions reached in the last two sections by
considering the simple example of massive gravity where we explicitly impose a
constraint on the system with the purpose to decrease the number degrees of
freedom in the system. However, contrary to our expectations we discover that
the previously \textquotedblleft silent\textquotedblright\ gravitational
degrees of freedom are forced by constraint to become propagating. This mimics
the situation with vanishing Hessian's determinant.

Let us consider a massive gravity with four scalars $\phi^{A}$ \ \cite{ChM1}
subject to the constraint
\begin{equation}
H=4,\qquad\Rightarrow\overline{h}=0,
\end{equation}
where $H^{AB}=g^{\mu\nu}\partial_{\mu}\phi^{A}\partial_{\nu}\phi^{B},$ and
$H=H^{AB}\eta_{AB}=4+\overline{h}.$

We will now examine whether the above constraint really reduces the original
six degrees of freedom to the five ones, required for the massive graviton.

We will analyze this system in different ways. First, we implement the
condition $\overline{h}=0$ in the action expanded to the second order in
perturbations%
\begin{equation}
S=\frac{1}{8}%
{\displaystyle\int}
d^{4}x\left[  \partial^{C}h_{B}^{A}\partial_{C}h_{A}^{B}-2\partial^{C}%
h_{C}^{A}\partial_{D}h_{A}^{\,\,D}+2\partial^{C}h_{C}^{A}\partial
_{A}h-\partial_{A}h\partial^{A}h-m^{2}\overline{h}_{B}^{A}\overline{h}_{A}%
^{B}\right]  .
\end{equation}
Taking into account that $\overline{h}_{AB}=h_{AB}+\partial_{A}\chi
_{B}+\partial_{B}\chi_{A}$ the last term here becomes%
\begin{equation}
-\frac{m^{2}}{8}\left(  h^{AB}h_{AB}+4\partial^{A}\chi^{B}h_{AB}+2\partial
^{A}\chi^{B}\partial_{A}\chi_{B}+2\partial^{A}\chi^{B}\partial_{B}\chi
_{A}\right)  .
\end{equation}
Using the constraint $\overline{h}=h+2\partial^{A}\chi_{A}=0$ we can write%
\begin{equation}
\chi_{A}=\widehat{\chi}_{A}-\frac{1}{2\partial^{2}}\partial_{A}h,\qquad
\end{equation}
where $\partial^{A}\widehat{\chi}_{A}=0.$

To determine the propagators, we fix the gauge by choosing the following gauge
conditions,%
\begin{align}
G  &  =H-4=0,\\
G^{A}  &  =\partial^{B}h_{B}^{A}-\frac{1}{2}\partial_{A}h-m^{2}\chi^{A}=0,
\end{align}
which have to be imposed via corresponding delta functions in the path
integral%
\begin{equation}%
{\displaystyle\int}
\left[  Dh^{\mu\nu}\right]  \left[  D\chi^{A}\right]  \delta\left(
\overline{h}\right)  \delta\left(  \partial^{B}h_{B}^{A}-\frac{1}{2}%
\partial^{A}h-m^{2}\chi^{A}\right)  e^{-S}.
\end{equation}
Next we add to the action the gauge fixing term%
\begin{equation}
\frac{1}{4\alpha}\left(  \partial_{A}h^{AB}-\frac{1}{2}\partial^{B}h-m^{2}%
\chi^{B}\right)  ^{2},
\end{equation}
and to simplify the expressions we set $\alpha=1$ to cancel the cross terms.
The action then reduces to:
\begin{align}
S  &  =\frac{1}{8}%
{\displaystyle\int}
d^{4}x\left[  \partial^{C}h_{B}^{A}\partial_{C}h_{A}^{B}-\frac{1}{2}%
\partial_{A}h\partial^{A}h-m^{2}h_{B}^{A}h_{A}^{B}\right. \nonumber\\
&  \left.  +2m^{4}\chi^{B}\chi_{B}-2m^{2}h\partial^{B}\chi_{B}-2m^{2}%
\partial^{A}\chi^{B}\partial_{A}\chi_{B}-2m^{2}\partial^{A}\chi^{B}%
\partial_{B}\chi_{A}\right] \nonumber\\
&  =\frac{1}{8}%
{\displaystyle\int}
d^{4}x\left(  -h_{B}^{A}\left(  \partial^{2}+m^{2}\right)  h_{A}^{B}+\frac
{1}{2}h\left(  \partial^{2}+m^{2}\right)  h+2m^{2}\chi^{B}\left(  \partial
^{2}+m^{2}\right)  \chi^{B}\right)  ,
\end{align}
where we have taken into account that $\overline{h}=h+2\partial^{A}\chi
_{A}=0.$ Substituting $\partial^{A}\chi_{A}=-\frac{1}{2}h$ \ gives%
\begin{equation}
2m^{2}\chi^{B}\left(  \partial^{2}+m^{2}\right)  \chi^{B}=2m^{2}\widehat{\chi
}^{B}\left(  \partial^{2}+m^{2}\right)  \widehat{\chi}^{B}-\frac{1}{2}h\left(
\partial^{2}+m^{2}\right)  \frac{m^{2}}{\partial^{2}}h,
\end{equation}
and finally the action reduces to
\begin{align}
S  &  =\frac{1}{8}%
{\displaystyle\int}
d^{4}x\left(  -h_{B}^{A}\left(  \partial^{2}+m^{2}\right)  h_{A}^{B}+\frac
{1}{2}h\left(  \partial^{2}+m^{2}\right)  \left(  1-\frac{m^{2}}{\partial^{2}%
}\right)  h\right. \nonumber\\
&  \qquad\qquad\left.  +2m^{2}\widehat{\chi}^{B}\left(  \partial^{2}%
+m^{2}\right)  \widehat{\chi}^{B}\right)  .
\end{align}
Isolating the traceless part of $h_{AB}$,
\begin{equation}
h_{AB}=h_{AB}^{T}+\frac{1}{4}\eta_{AB}h,
\end{equation}
we obtain for the trace part the following action
\begin{align}
&  -\frac{1}{4}h\left(  \partial^{2}+m^{2}\right)  h+\frac{1}{2}h\left(
\partial^{2}+m^{2}\right)  h-\frac{1}{2}h\left(  \partial^{2}+m^{2}\right)
\frac{m^{2}}{\partial^{2}}h\nonumber\\
&  =\frac{1}{4}h\left(  \partial^{2}+m^{2}\right)  \left(  1-\frac{2m^{2}%
}{\partial^{2}}\right)  h, \label{hpart}%
\end{align}
which clearly contains a tachyonic mode of mass $-2m^{2}.$ In fact, the scalar
mode has a propagator
\begin{equation}
\frac{p^{2}}{\left(  p^{2}-m^{2}\right)  \left(  p^{2}+2m^{2}\right)  }%
=\frac{1}{3}\left(  \frac{1}{p^{2}-m^{2}}+\frac{2}{p^{2}+2m^{2}}\right)  ,
\end{equation}
which represents a combination of a physical spin zero state of mass $m$ and a
spin zero \textit{tachyonic state} of mass $m\sqrt{2}.$ Together with two
degrees of freedom of originally massless graviton and two degrees of freedom
induced by a vector part of scalar fields this makes six degrees of freedom in
total. Thus the imposed constraint did not reduce the number of the original
degrees of freedom but, instead of that, forced the originally
\textquotedblleft silent\textquotedblright\ metric components to propagate.

Now we will show how to arrive to the same conclusion by enforcing the
constraint $\overline{h}=0$ through a Lagrange multiplier in the action%
\begin{equation}
S=-\frac{1}{2}%
{\displaystyle\int}
d^{4}x\sqrt{g}R\left(  g\right)  +\frac{m^{2}}{8}%
{\displaystyle\int}
d^{4}x\sqrt{g}\left(  \ 4-H_{AB}H^{AB}+\lambda\left(  H-4\right)  \right)  .
\end{equation}
Varying with respect to $\delta g^{\mu\nu}$ we obtain
\begin{align}
\frac{1}{2}\left(  R_{\mu\nu}-\frac{1}{2}g_{\mu\nu}R\right)   &  =-\frac
{m^{2}}{16}g_{\mu\nu}\left(  4-H_{AB}H^{AB}+\lambda(H-4)\right)  \nonumber\\
&  -\frac{m^{2}}{4}H^{AB}\partial_{\mu}\phi_{A}\partial_{\nu}\phi_{B}%
+\frac{m^{2}}{8}\partial_{\mu}\phi^{A}\partial_{\nu}\phi^{A}\lambda,
\end{align}
while the $\lambda$ equation gives
\begin{equation}
H=4,
\end{equation}
and the $\phi^{A}$ equations are
\begin{equation}
\frac{2m^{2}}{\sqrt{g}}\partial_{\mu}\left[  \sqrt{g}\left(  -2g^{\mu\nu
}\partial_{\nu}\phi^{B}H_{AB}+g^{\mu\nu}\partial_{\nu}\phi^{A}\lambda\right)
\right]  =0.
\end{equation}
In the linearized approximation (with $\lambda=2+\overline{\lambda}$ to cancel
zero order term) the above two equations become:
\begin{align}
0 &  =\left(  \partial^{2}+m^{2}\right)  h_{AB}-\partial_{A}\partial^{C}%
h_{CB}-\partial_{B}\partial^{C}h_{AC}\nonumber\\
+ &  \eta_{AB}\left(  \partial^{C}\partial^{D}h_{CD}-\partial^{2}h\right)
+m^{2}\left(  \partial_{A}\chi_{B}+\partial_{B}\chi_{A}\right)  -\frac{1}%
{2}m^{2}\eta_{AB}\overline{\lambda},\label{curvature}%
\end{align}
and%
\begin{align}
&  0=\partial_{\mu}\left[  \left(  1-\frac{h}{2}\right)  \left(  -2\eta
^{\mu\nu}-2h^{\mu\nu}\right)  \left(  \delta_{\nu}^{B}+\partial_{\nu}\chi
^{B}\right)  \left(  \eta_{AB}+\overline{h}_{AB}\right)  \right.  \nonumber\\
&  \qquad\left.  +\left(  1-\frac{h}{2}\right)  \left(  \eta^{\mu\nu}%
+h^{\mu\nu}\right)  \left(  \delta_{\nu}^{A}+\partial_{\nu}\chi^{A}\right)
\left(  2+\overline{\lambda}\right)  \right]  .\label{linearized}%
\end{align}
Equation (\ref{linearized}) simplifies to
\begin{equation}
\partial_{A}\overline{\lambda}-2\partial^{\mu}\overline{h}_{\mu A}=0.
\end{equation}
Using the gauge fixing condition
\begin{equation}
\partial^{B}h_{B}^{A}-\frac{1}{2}\partial_{A}h-m^{2}\chi^{A}=0,
\end{equation}
we obtain
\begin{align}
\partial_{A}\overline{\lambda} &  =2\partial_{\mu}\left(  h_{\mu A}%
+\partial_{\mu}\chi_{A}+\partial_{A}\chi_{\mu}\right)  \nonumber\\
&  =2\left(  \partial^{2}+m^{2}\right)  \chi_{A}.\label{previous}%
\end{align}
On the other hand, the curvature equation (\ref{curvature}) taking into
account the gauge condition, gives
\begin{equation}
\left(  \partial^{2}+m^{2}\right)  \left(  h_{AB}-\frac{1}{2}\eta
_{AB}h\right)  =\frac{1}{2}m^{2}\eta_{AB}\overline{\lambda},\label{above}%
\end{equation}
which is consistent with the divergence of equation (\ref{previous}). Trace of
equation (\ref{above}) leads to
\begin{equation}
-\left(  \partial^{2}+m^{2}\right)  h=2m^{2}\overline{\lambda}.
\end{equation}
The $\chi_{A\text{ }}$equation can be written as
\begin{equation}
\left(  \partial^{2}+m^{2}\right)  \left(  \widehat{\chi}_{A}-\frac{1}{2}%
\frac{\partial_{A}}{\partial^{2}}h\right)  =\frac{1}{2}\partial_{A}%
\overline{\lambda},
\end{equation}
which implies that
\begin{align}
\left(  \partial^{2}+m^{2}\right)  \widehat{\chi}_{A} &  =0,\nonumber\\
\overline{\lambda} &  =-\frac{\partial^{2}+m^{2}}{\partial^{2}}h=-\frac
{1}{2m^{2}}\left(  \partial^{2}+m^{2}\right)  h.
\end{align}
Thus
\begin{equation}
\left(  \partial^{2}+m^{2}\right)  \left(  1-\frac{2m^{2}}{\partial^{2}%
}\right)  h=0,
\end{equation}
which is the same as equation of motion derived from equation (\ref{hpart}).
Thus we confirm the previous result.

Finally and most directly, the tachyonic mode can be seen if we consider the
equations of motion of linearized massive gravity \cite{ChM1}, taking into
account that
\begin{equation}
\overline{h}=0,\label{constraint}%
\end{equation}
which then take the following form
\begin{equation}
\partial^{2}\overline{h}_{\nu}^{\mu}-\partial_{\nu}\partial^{\rho}\overline
{h}_{\rho}^{\mu}-\partial^{\mu}\partial_{\rho}\overline{h}_{\nu}^{\rho}%
+\frac{1}{2}\delta_{\nu}^{\mu}\partial^{\rho}\partial_{\sigma}\overline
{h}_{\rho}^{\sigma}+m^{2}\overline{h}_{\nu}^{\mu}=0\label{pauli}%
\end{equation}
Substituting into these equations the following ansatz
\begin{align}
\overline{h}_{i}^{0} &  =\frac{2}{3m^{2}+2\Delta}\partial_{i}\partial
^{0}\overline{h}_{0}^{0}\\
\overline{h}_{j}^{i} &  =-\frac{\left(  \delta_{j}^{i}m^{2}+2\partial
_{j}\partial_{i}\right)  }{3m^{2}+2\Delta}\overline{h}_{0}^{0}%
\end{align}
one can verify that equations $\left(  \ref{constraint}\right)  $ and $\left(
\ref{pauli}\right)  $ are both satisfied provided that
\begin{equation}
\left(  \partial_{0}^{2}-\Delta-2m^{2}\right)  \overline{h}_{0}^{0}=0
\end{equation}
One can check that all other components of $\overline{h}_{\nu}^{\mu}$ satisfy
the same equation, implying that the tachyonic mode with mass $-2m^{2}$ is propagating.

\section{Conclusions}

In this paper we have studied the action for massive gravity which is claimed
to be ghost free \cite{HR} \cite{GR2}. We have shown that in this theory the
Hessian for the scalar fields, used to generate the graviton mass, has a
vanishing determinant, thus implying that at least one of them can be
eliminated. However, this does not reduce the number of degrees of freedom of
the whole system because the originally silent component of the metric starts
to propagate. We have argued then that this seems is a general behavior of
massive gravity, whenever a constraint is imposed to reduce the number of
degrees of freedom.

\begin{acknowledgement}
The work of AHC is supported in part by the National Science Foundation
\ Phys-0854779, and Phys-1202671. The work of VM is supported by
\textquotedblleft Chaire Internationale de Recherche Blaise Pascal
financ\'{e}e par l'Etat et la R\'{e}gion d'Ile-de-France, g\'{e}r\'{e}e par la
Fondation de l'Ecole Normale Sup\'{e}rieure\textquotedblright, by TRR 33
\textquotedblleft The Dark Universe\textquotedblright\ and the Cluster of
Excellence EXC 153 \textquotedblleft Origin and Structure of the
Universe\textquotedblright. \bigskip
\end{acknowledgement}

\end{document}